\documentstyle[12pt,epsfig]{article}

\large
\newcommand{\be}{\begin{eqnarray}}
\newcommand{\ee}{\end{eqnarray}}
\newcommand{\nn}{\nonumber}
\newcommand{\e}{{\rm e}}

\begin{document}
\setlength{\baselineskip}{17pt}
\pagestyle{empty}
\vfill 
\eject
\begin{flushright}
SUNY-NTG-2002/06
\end{flushright}

\vskip 2.0cm
\centerline{\Large \bf
Critical statistics for non-Hermitian matrices}
\vskip 0.4cm
\centerline{\Large \bf }

\vskip 1.2cm
\centerline{
A.M. Garc\'{\i}a-Garc\'{\i}a${}^{1}$,
S.M. Nishigaki${}^{2}$
and 
J.J.M. Verbaarschot%
\footnote{2002 James H. Simons Fellow}${}^{1}$}

\vskip 0.2cm

\centerline{${}^{1}$\it
Department of Physics and Astronomy, SUNY,
Stony Brook, New York 11794-3800}
\centerline{${}^{2}$\it
Department of Physics, University of Connecticut,
Storrs, Connecticut 06269-3046}
\vskip 1.5cm

\centerline{\bf Abstract}

\noindent{We introduce a
 generalized ensemble of nonhermitian matrices
interpolating between
 the Gaussian Unitary Ensemble, the Ginibre ensemble and the Poisson ensemble.
The joint eigenvalue distribution of this model is obtained by means
of an extension of  the Itzykson-Zuber formula to general complex
matrices. Its correlation functions  are studied both in the
case of weak nonhermiticity and in the case of strong nonhermiticity.
In the weak nonhermiticity
limit we show that the spectral correlations in the bulk of the spectrum
display critical statistics: the asymptotic linear behavior of the
number variance is already approached 
for  energy differences of the order
of the eigenvalue spacing. To lowest order, its slope does not
depend on the degree of nonhermiticity.
Close the edge, the spectral correlations are similar
to the Hermitian case.
In the strong nonhermiticity limit the crossover behavior from the Ginibre
ensemble to the Poisson ensemble first appears
close to the surface of the spectrum.
Our model may be relevant for the description of  the  spectral
correlations of an open disordered system close to an Anderson transition.


\vfill
\noindent

\eject
\pagestyle{plain}
\section{Introduction}
\setcounter{page}{1}

Nonhermitian Random Matrix Models were first introduced by Ginibre in 1965
\cite{Ginibre}. His motivation was to describe the statistical properties
of nuclear resonances with a finite width in complete analogy with the
description of the position of resonances by means
Hermitian Random Matrix Ensembles as introduced by Wigner and Dyson
\cite{Dyson-Mehtabook}. Since then,
eigenvalues of nonhermitian operators occurring in many different fields
have been analyzed in
terms of nonhermitian random matrix models, usually with additional
ingredients.
We mention several examples.
The statistical properties of the
poles of $S$-matrices have been analyzed in great detail in
\cite{VWZ,Essen,john}.
In QCD, the Euclidean Dirac operator in QCD at nonzero chemical potential
 (which can be interpreted as an imaginary vector potential),
  is nonhermitian resulting in the failure of the quenched
approximation \cite{everybody}.
Both this failure and the generic properties of the complex
Dirac spectrum
have been explained fully in terms of a
nonhermitian Random Matrix Model with the global symmetries of QCD
\cite{misha,muthree,wet,TV,gernot}.
 Recently, a delocalization transition was found
in a one-dimensional lattice model with an imaginary vector potential
\cite{hatano,efetov}.
Statistical correlations predicted by the Ginibre
ensemble have been found
dissipative quantum maps \cite{haa,haa1,haa2}. Eigenvalue
spacings of the
Floquet matrix of a Fokker Planck equation have been described
in terms of Ginibre statistics \cite{reich}.
In \cite{som,som2} an ensemble of asymmetric real matrices,
closely related to the Ginibre ensemble,
was utilized to model the dynamics of a neural network.

Among more  mathematically oriented  works we mention the exact
calculation of the correlation functions of an ensemble of
normal random matrices with an  arbitrary polynomial probability potential
\cite{yu,oas}. Nonhermitian ensembles have been analyzed in terms of
associated hermitian ensembles \cite{girko,zee}.
Correlations of eigenfunctions have been studied in
the Ginibre ensemble \cite{eigenfunctions}.
Another intriguing application is
the  description of an analytic curve by the
boundary of the support of the complex spectrum of a nonhermitian
Random Matrix Theory \cite{zeespect,Kostov}.  Finally, we point out that there are
interesting relations between the eigenvalues of complex
matrices and the positions of particles in certain
two dimensional physical systems \cite{forrester-98,forrester,hastings}.
For example, the Ginibre model
is equivalent to a Coulomb problem in two dimensions \cite{Ginibre}.

Based on the magnitude of the imaginary part of the eigenvalues
we distinguish two types of nonhermiticity:
weak nonhermiticity and strong nonhermiticity.
Weak nonhermiticity is the limit of large matrices
when the imaginary part of the eigenvalues
remains comparable with the mean separation of eigenvalues along the real
axis.
This limit was identified in \cite{Fyodorov, Fyodorov1,Fyodorov2},
but was used earlier in  the statistical theory of  $S$-matrices \cite{VWZ}.
Strong nonhermiticity refers to cases for which the real
and imaginary parts of the eigenvalues remain of the same order of magnitude
in the thermodynamic limit. In this article we consider both types of
nonhermiticity.

An important concept in the understanding of disordered systems is the
Thouless energy. We will define this energy scale as the
energy difference below which
the eigenvalues are correlated according to Random Matrix Theory.
In diffusive disordered systems, in the thermodynamic
limit, both the eigenvalue spacing and the Thouless energy
approach zero
whereas the number of eigenvalues in between them approaches
infinity.
In this article we will consider critical statistics
\cite{shk,log2,bleck,kra1}
which refers to the case that
the ratio of the
Thouless energy and the eigenvalue spacing remains finite in
the thermodynamic limit.
A Hermitian Random Matrix
model for critical statistics
was proposed in \cite{MSN}. In that model the correlations of the
 eigenvalues decay exponentially beyond a Thouless energy resulting in an
asymptotically  linear behavior of the number variance with
slope (level compressibility) less than one.
In this article we generalize this model
to complex eigenvalues  and analyze its properties.  In the Ginibre model
the two-point correlation
function of eigenvalues in the bulk of the spectrum
drops off exponentially on the scale of the distance between
the eigenvalues. It is therefore no surprise that
we will find the same bulk correlations in such generalized Ginibre model.
However, we find nontrivial long range surface correlations,
characteristic of a two-dimensional Coulomb liquid.
In the case of weak nonhermiticity we expect to find critical
statistics similar to the Hermitian model. The analysis of this case
is the main objective of this article.

Critical statistics is associated with the multifractal behavior of
the eigenfunctions \cite{kra1,quezada,mirlin}.
The critical Hermitian model
introduced in {\cite{MSN}}
has the unitary invariance of the Gaussian Unitary Ensemble
with eigenvectors that are distributed according to the measure of
the unitary group. This is no  contradiction:
multifractality of wave functions
occurs in a specific basis in which disorder competes with a hopping term.
Indeed, in \cite{chalker,chalker2}  it was found
that the fractal dimension of the wave function determines
the asymptotic slope of the number variance.

Among others, critical statistics
have been utilized  to describe the spectral correlations
 of disordered system at the Anderson transition in three dimensions
{\cite{shk,nishigaki}, two dimensional Dirac fermions
in a random potential {\cite{caux}}, quantum Hall transition
\cite{klesse} and QCD Dirac operator in a liquid of instantons
\cite{james,us}.
The scope of universality of critical statistics
is still under debate.

Our Random Matrix Model is introduced in section 2. The case or strong
nonhermiticity and weak nonhermiticity are analyzed in sections 3 and
4, respectively. Among
others we derive a closed expression for
 the two-point correlation function in both limits. Results for
the number variance are discussed in section 5
and  concluding remarks are macritical statistics
is still under debate.de in section 6.

\section{Introduction of the model}

Recently, a Hermitian random matrix model
for critical statistics was introduced by
Moshe, Neuberger and Shapiro \cite{MSN}. This model, which
interpolates
between Wigner-Dyson statistics and Poisson statistics, is
defined by the joint eigenvalue probability distribution
\be
P(H) dH = dH \int dU  \e^{-(1+b){\rm Tr}\, H^2 +
b {\rm Tr}\, UHU^\dagger H^\dagger},
\label{mns-model}
\ee
where $H$ is a Hermitian $n\times n$ matrix. The integral is
over the unitary group with invariant measure denoted by $dU$.
Critical statistics \cite{kra1} is obtained in the  thermodynamic limit
with $b$ scaling as  $b= h^2 n^2$ at fixed $h$ . In that case, the
two-point correlation
function decays exponentially at large distances
and the number variance has an asymptotic linear behavior
with slope less than one.
In the thermodynamic limit, Wigner-Dyson
statistics is obtained for a weaker $n$-dependence of $b$, and Poisson
statistics is found for a stronger $n$-dependence of $b$.

In this article we are interested in ensembles of nonhermitian random
matrices.  The study of random matrices with no restrictions imposed
was initiated by the classical work of Ginibre \cite{Ginibre}.
He found closed expressions
for the two-point correlation function of the eigenvalues of a Gaussian
ensemble of random matrices with complex entries.

An ensemble that interpolates between the Ginibre ensemble
and the Wigner-Dyson ensemble of Hermitian matrices was introduced in
\cite{Fyodorov,Fyodorov1}
 \be
P(C) dC \sim dC\,
\e^{-\frac{1}{1-\tau^2}{\rm Tr } C^ \dagger C
+ \frac{\tau}{2(1-\tau^2)}  {\rm Tr}\,(C^2 + (C^\dagger)^2)} .
\ee
Here, $C$ is an arbitrary $n\times n$
complex matrix with integration measure given
by the product of the real and imaginary parts of the differentials of
the matrix elements of $C$. For $\tau =0 $ this model reduces to
the Ginibre ensemble whereas for $\tau = 1$ ($-1$) it reduces to a
Gaussian ensemble of (anti-)Hermitian matrices. The eigenvalues of this
ensemble are scattered inside an ellipse with eccentricity given by
$2\sqrt \tau/(1+\tau)$.

The joint eigenvalue distribution can be obtained by using two
alternative decompositions
\be
C= U T U^\dagger \qquad {\rm and} \qquad C = V \Lambda V^{-1},
\label{decomp}
\ee
where $U$ is a unitary matrix, $V$ is a similarity transformation, $T$
a upper triangular matrix and $\Lambda $ a diagonal matrix.
The diagonal
matrix elements of $T$ coincide with the  complex eigenvalues
$\Lambda_{kk}=z_{k}$.
The invariant measure factorizes as \cite{Dyson-Mehtabook}
\be
dC \sim   dU dT \Delta (\{\Lambda_{kk}\}) \Delta(\{\Lambda^*_{kk}\})
\ee
with the Vandermonde determinant defined by
\be
\Delta(\{z_k\}) =\prod_{k<l}^n (z_k - z_l).
\label{vandermonde}
\ee
Since the Gaussian integral over the off-diagonal matrix elements of $T$
factorizes it can be performed trivially.
The integral over $U$ is equal to the group
volume. The joint probability distribution of the eigenvalues is thus
given
by
\be
P(\Lambda) d\Lambda \sim d\Lambda\, |\Delta(\Lambda)|^2\,
\e^{-\frac{1}{1-\tau^2}\sum_{i=0}^n [|z_i|^2 -
\frac{\tau}{2} (z_i^2 + (z_i^*)^2) ]}.
\label{yan-model}
\ee
This model has been analyzed in two domains: weak nonhermiticity and
strong nonhermiticity. In the first case the thermodynamic
limit is taken at fixed $n(1-\tau)$, whereas in the case of strong
nonhermiticity $-1 <\tau < 1$ remains fixed for $n\to \infty$.
The two-point correlation function of this model was derived in
\cite{Fyodorov}.

In this article, we analyze a model that interpolates in between
the models defined in eqs. (\ref{mns-model}) and (\ref{yan-model}).
Our random matrix model is defined by
\be
P(C) dC \sim dC\, \e^{-a_{1}{\rm Tr}\, C^ \dagger C
- \frac {a_2}2  {\rm Tr}\,(C^2 + (C^\dagger)^2)}
\int dU \e^{a_3 {\rm Tr}\, U C U^\dagger C^\dagger}.
\label{our-model}
\ee
where is $C$  an arbitrary complex $n\times n$ matrix,
and $dU$ is the Haar measure of the unitary group ${\rm U}(n)$. In the
special case of $C$
being  a normal matrix ($[C,C^\dagger ] = 0$), a
unitary transformation brings $C$ to a diagonal form  and
the integral over $U$ is the standard  Itzykson-Zuber integral
\cite{zuber} given by
\be
\int dU\,\e^{a_3 {\rm Tr}\, U C U^\dagger C^\dagger} = \frac{\det
\e^{a_3 z_i z_j^*}}{\Delta(\{z_k\}) \Delta(\{z^*_k\})},
\label{IZ}
\ee
where the  $z_i$ are the eigenvalues of $C$. One thus
finds the joint eigenvalue distribution
\be
P(\{z_k\}) \sim
\e^{-
\sum_{i=n}^n    \left [ a_{1}|z_i|^2 + \frac {a_2} 2 (z_i^2 +z_i^{*2})
\right ]}
{\rm det }\left [ \e^{a_3 z_i z_j^* }\right ].
\ee
In the next paragraph we will show that
this result is valid even if $C$ is an arbitrary complex matrix that
can be decomposed  according to (\ref{decomp}).

We start from the triangular decomposition
$C = U T U^\dagger$. Since $T$ is an upper-triangular matrix,
the exponent in the integral over $U$ in
(\ref{our-model}) is then given by
\be
{\rm Tr}\, U C U^\dagger C^\dagger  =\sum_{{j \le k \atop i \le l}}~
U_{ij} T_{jk} U_{lk}^*T^*_{il}.
\label{expt}
\ee
After performing
a trivial ${\rm U}(1)$ integration,
the integral over $U$ in (\ref{IZ}) is over ${\rm SU}(n)$.
The generating function for such integrals
is given by
\be
\int_{U\in {\rm SU}(n)} \!\!\!\!
dU\, \e^{{\rm Tr}\,( J U^\dagger  + J^\dagger U)} =
F(\det J, \det J^\dagger,  \{ {\rm Tr}\, (J^\dagger J)^k \}),
\label{gen}
\ee
where $J$ is a complex $n\times n$ matrix and the functional form
of the r.h.s., with $k$ running  over all positive integers,
follows from the invariance of the group integral. In the expansion
of the exponent (\ref{expt}) all terms have the same number of
factors $U$ and $U^*$. By differentiating (\ref{gen}) with respect to
$J$ and $J^*$ at $J=0$, we find that such terms can be only non
vanishing
if the sum of the indices of $U$ is equal to the
sum of the indices of $U^*$ (for the terms that enter in the expansion
of the determinant the sum of the first indices is equal to to
sum of the second indices).
We thus find that in the expansion of (\ref{expt})  all terms with
off-diagonal elements of $T$ or $T^\dagger$ vanish after integration.
We conclude that the result (\ref{IZ}) for the Itzykson-Zuber integral
is also valid for
an arbitrary complex matrix $C$ with eigenvalues $z_k$.

For convenience, the constants in the joint eigenvalue distribution of
(\ref{our-model}) will be parameterized as
\be
a_1 &=& \frac{\lambda}{1-\tau^2},\nn \\
a_2 &=& -\frac{\lambda \tau }{1-\tau^2} +
\frac{\lambda \alpha^2}{\tau (1-\alpha^2)},\nn \\
a_3 &=&\frac{\lambda \alpha}{\tau (1-\alpha^2)}.
\ee
After a rescaling of the matrix elements of $C$ by a factor $1/\sqrt
\lambda$
the joint eigenvalue distribution of the model (\ref{our-model})
reduces to
\be
P(\Lambda) d\Lambda \sim d\Lambda\,
\e^{ -\sum_{i=1}^n \left[\frac{1}{1-\tau^2}|z_i|^2 -
\frac{\tau }{2(1-\tau^2)} (z_i^2 +z_i^{* 2})
+\frac{\alpha^2}{2\tau (1-\alpha^2)}(z_i^2+z_i^{*2}) \right ] }
{\rm det }\left [ \e^{ \frac{\alpha}{\tau (1-\alpha^2)} z_i z_j^*}
\right ].
\label{our-joint}
\ee
We will analyze this model in two limits.
The case when $1-\tau$ remains finite
in the thermodynamic limit will
be referred to as strong nonhermiticity.
In this class of models we will consider the limiting case
of zero eccentricity
\be
\alpha \to 0, \quad \tau \to 0 \quad
{\rm with} \quad \frac \alpha \tau
=b\quad \mbox{fixed},
\label{snh}
\ee
which reduces to the Ginibre model in the limit in which
the parameter $b$ is taken to zero.
On the other hand, the case of
weak nonhermiticity \cite{Fyodorov} is defined by the limit
\be
\tau\to 1,\ n\to\infty,\quad
(1-\tau) n = a^2 \quad \mbox{fixed}.
\label{wnh}
\ee

Finally, let us mention that the wave functions of our model are
distributed
according to the invariant Haar measure of ${\rm U}(n)$.
It could be that
for diagonal $U$ in (\ref{our-model}) the wave functions show a
multifractal
behavior, but that this property is obscured by averaging over all $U$
whereas eigenvalue correlations remain unaffected.






\section{Strong nonhermiticity}

In this section we consider the case of strong nonhermiticity
(\ref{snh}).
In order to rewrite the Itzykson-Zuber determinant in
Eq.(\ref{our-joint})
in terms of an expectation value of two Slater determinants,
we expand the exponential as
\be
 \e^{b z_i z_j^*} &=&  \sum_{m=0}^\infty \frac {b^m}{m!} z_i^m
(z_j^*)^m\label{istrong} .
\ee
By a series of elementary manipulations we find
\be
\det \e^{b z_i z_j^*}
&=&\sum_{m_1=0}^\infty \cdots \sum_{m_n=0}^\infty
\frac {b^{m_1+\dots+ m_n}}{m_1!\cdots m_n!}
\sum_{\pi\in {\rm S}_n}
(-1)^{\sigma(\pi)} z_1^{m_1}(z_{\pi(1)}^*)^{m_1}\cdots
z_n^{m_n}(z_{\pi(n)}^*)^{m_n}\nn \\
&=& \sum_{m_1< m_2 < \cdots < m_n}
b^{m_1+\dots +m_n}
\det \frac{z_i^{m_j}}{\sqrt{m_j!}}
\det \frac{z_k^{* m_l}}{\sqrt{m_l!}}. \label{manipul}
\ee
Including the other factors of the joint probability distribution
we thus find
\be
P(z) dz \sim
 \sum_{m_1< m_2 < \cdots < m_n}
  b^{m_1+\dots +m_n}
\det \phi_{m_j}(z_i) \det \phi_{m_l}(z_k^*),
\ee
where the normalized wave functions are given by
\be
\phi_{k}(z) = {\frac{1}{\sqrt{\pi k!}}} z^k \e^{-|z|^2/2}
\label{orthogonality}
\ee
satisfy the orthogonality relation
\be
\int d^2 z \,\phi_k^*(z)\phi_l(z) = \delta_{kl}.
\ee
They are the single particle wave functions of the lowest Landau level
of
a particle with unit mass
in a constant magnetic field perpendicular to the plane. The
Hamiltonian of this system is given by
$(z=x+i y)$
\be
H = \frac 12 (i\partial_x -  y)^2 + \frac 12(i\partial_y + x )^2 .
\label{hamiltonian}
\ee
and the  corresponding Schr\"odinger equation reads
\be
H \phi_k(z) = \phi_k(z).
\label{schrodinger}
\ee
If we write
\be
b = \e^{-\beta}
\ee
the joint probability distribution is equal to the diagonal
element of the $n$-body density matrix
of the lowest Landau level fermions  at temperature
$1/\beta$, with an additional
degeneracy-breaking Hamiltonian given by the absolute value
of the angular momentum
\be
L=iy\partial_x  - ix\partial_y ,
\ee
or equivalently of 
\be
\tilde{H}=H+2L=
\frac 12 (i\partial_x +  y)^2 + \frac 12(i\partial_y - x )^2 .
\label{schrm}
\ee
The average spectral density
$\rho_n(z)$, which can be interpreted as the
one-particle density, is obtained by integrating the joint eigenvalue
density over all coordinates except one. By using the orthogonality
relations (\ref{orthogonality}) one easily finds
\be
\rho_n(z) = 
\frac{1}{Z_n}
\sum_{m_1< m_2 < \cdots < m_n} \sum_{i=1}^n
\e^{-\beta(m_1 + \cdots + m_n)}
\phi_{m_i}(z) \phi_{m_i}(z^*),
\ee
or in an occupation number representation
\be
\rho_n(z) = \frac{1}{Z_n}
\sum_{n_1+ n_2 + \cdots = n}
 \e^{-\beta\sum_p p  n_p}
 \sum_{k=0}^\infty n_k
\phi_{k}(z) \phi_{k}(z^*),
\ee
where the occupation number $n_k$ runs over $\{0, 1\}$.
The partition function $Z_n$ is defined in the usual way
\be
Z_n =\sum_{n_1+ n_2 + \cdots = n}
 \e^{-\beta \sum_p p  n_p}.
\ee
Such sums can be easily evaluated in the grand canonical ensemble
\be
\rho(z) &=& \frac 1Z \sum_n \zeta^n {Z_n}
\rho_n(z) \nn \\ &=&
\sum_{k=1}^n
\frac{\phi_{k}(z) \phi_{k}(z^*)}{1 + \zeta^{-1} \e^{\beta k}}
\equiv \frac{1}\pi k(z,z),
\ee
where we have introduced the prekernel
\be
k(z_1, z_2) = \e^{-z_1 z_2^*} \sum_{k=0}^\infty \frac{(z_1 z_2^*)^k}
{k!( 1 + \zeta^{-1} \e^{\beta k} )}.
\label{prekernel}
\ee
The fugacity $\zeta$ is determined by the normalization of the
one-particle density
\be
n = \sum_{k=0}^\infty
\frac{1}{1 + \zeta^{-1} \e^{\beta  k}} .
\ee
For $\beta \ll 1$ the sum can be converted into an integral
resulting in
\be
\zeta = \e^{n\beta } -1.
\label{zeta}
\ee

Similarly, the two-point correlation function is obtained by
integrating over all
eigenvalues except two. Again by going to the grand canonical ensemble
one easily derives that the connected
two-point correlation can be factorized in the
result for the Ginibre ensemble and the prekernel (\ref{prekernel})
\be
R_2(z_1,z_2) = -\frac{1}{\pi^2} \e^{-|z_1-z_2|^2} |k(z_1,z_2)|^2.
\label{r2strong}
\ee

For $\beta \ll 1$ but $n \beta \gg 1$ a partial resummation of
the prekernel (\ref{prekernel}) results in
\be
k(z_1, z_2) = \sum_{k=0}^\infty \frac{\Gamma(k+1,z_1 z_2^*)}
{k!} \frac{\beta }
{4\cosh^2(\beta (k-n)/2)},
\ee
where $\Gamma(k,x)=\int_{x}^{\infty}t^{k-1}\e^{-t}dt$
is the incomplete $\Gamma$-function. For $\beta \to 0$
it is justified to make the approximation
\be
\frac{1}{1 + \e^{\beta (k-n)}} -
\frac{1}{1 + \e^{\beta (k+1-n)}}
\approx \frac{\beta }
{4\cosh^2(\beta (k-n)/2)}.
\ee
In the remainder of this subsection we will evaluate the
prekernel in several  limiting situations.

If the distance of $z_1$ and $z_2$
(both inside the disk of eigenvalues)
to the surface of the disk is much
larger than $\beta $, the numerator attains its maximum value when
the Fermi-Dirac factor is close to unity. In that case the
Fermi-Dirac
distribution can be replaced by a sharp cutoff and the two-point
correlation function is given by
\be
R_2(z_1,z_2) = -\frac{1}{\pi^2} \e^{- |z_1- z_2|^2}.
\ee
Inside the disk the average spectral density is $1/\pi$. The unfolded
two-point spectral correlation function thus coincides with the Ginibre
result.

A more interesting situation arises in
case both $z_1$ and $z_2$ are close to the surface of the disk of
eigenvalues.
A nontrivial thermodynamic limit of the surface correlations is obtained
for
\be
\beta &\sim &\frac{1}{\sqrt n},\nn \\
|z_1 z_2^*| &\sim& n ,\nn \\
{\rm arg}(z_1 z_2^*) &\sim& \frac{1}{\sqrt{n}}.
\ee
Using the asymptotic expansion for the incomplete $\Gamma$-function
 we find
\be
k(z_1, z_2) &=& \frac{\beta }{\sqrt \pi} \sum_{k=0}^\infty
\frac{{\rm Erfc}((z_1z_2^*-k)/\sqrt{2k})}
{4 \cosh^2(\beta (k-n)/2)}\nn \\
&\approx & \frac{\beta }{\sqrt \pi} \int_{-\infty}^\infty dt
\frac{{\rm Erfc}((z_1z_2^*-n -t)/\sqrt{2(n+t)})}
{4 \cosh^2(\beta t /2)} ,
\label{ints}
\ee
where ${\rm Erfc}(x)=\int_{x}^{\infty}\e^{-t^2}dt$.
We parameterize the vicinity of the surface of the domain of
eigenvalues
as
\be
z_k =\sqrt{n}+s_k, \quad k = 1, 2,
\qquad {\rm and}\qquad
 s = \frac{s_1+s_2^*}{2},
\ee
where $n \gg 1$ and
$|s_k| \ll \sqrt{n}$.
Introducing  the scaled temperature $h$ by
\be
\beta =\frac{1}{h\sqrt{n}},
\label{hn}
\ee
the prekernel simplifies  for $n \rightarrow \infty$ to
\be
k(z_1,z_2) =
\frac{2}{\sqrt{\pi}}\int_{-\infty}^{\infty}dt\frac{{\rm Erfc}(\sqrt{2}
(s-ht))}{4\cosh^2 t} .
\ee
To the leading order in $h$, this expression
can be simplified further,
\be
k(z_1,z_2) &=&
\frac{2}{\sqrt{\pi}}\int_{s\sqrt 2}^\infty dy\, \e^{-y^2}
\int_{-\infty}^{\infty}dt\frac{\e^{2\sqrt 2\, y ht}}
{4\cosh^2 t}\nn \\
&=&
 2\sqrt{2{\pi}} \int_{s}^\infty dy \frac {y\,\e^{-2y^2}
 }{\sin(2\pi y h)}.
\ee
For $s \gg 1$, the above integral dominated by the lower end point
and is approximated by
\be
k(z_1,z_2) &\sim&
\sqrt{\frac{\pi}2} \frac {h\, \e^{-2s^2} }{\sin(2\pi s h)}.
\ee
Accordingly, the spectral density near
the edge to the leading order in $h$ is  given by
\be
 \rho(z=\sqrt n + s)  &=& \frac{1}\pi k(z,z)~=~
\frac{2\sqrt{2}}{\sqrt {\pi}}
\int_{s}^\infty dy \frac {y h\, \e^{-2y^2} }{\sin(2\pi y h)}\nn \\
&\sim &
\frac{1}{\sqrt {2\pi}} \frac{h\,\e^{-2s^2}}{\sin(2\pi s h)}.
\ee
At the zero temperature $h\to 0$, it reduces to
the spectral density for the Ginibre ensemble
close to the edge given by \cite{Dyson-Mehtabook}
$ \rho(s) ={\e^{-2s^2}}/(2\pi)^{3/2}s$.
Likewise, the two-point function given by (\ref{r2strong})
simplifies to
\be
R_2(z_1=\sqrt n +s_1, z_2=\sqrt n + s_2) =
-\frac{1}{2\pi} \frac {h^2 \e^{-[(s_1+s_1^*)^2+(s_2+s_2^*)^2] /2}
}{|\sin(\pi (s_1+s_2^*) h)|^2}.
\ee
 for $|s_1+s_2^*| \gg 1$.
As a consistency check, we find that  the zero temperature limit for
$y_1-y_2 \gg x_k$ (with $s_k = x_k + i y_k$) ,
\be
R_2(z_1,z_2) = - \frac{1}{2\pi^3}
\frac{\e^{-2(x_1^2+x_2^2)}}{(y_1-y_2)^2}.
\label{surfacecor}
\ee
is in agreement with the result in \cite{jancovici-95} although
different prefactors have appeared in the literature
\cite{smith-82,forrester-98}. We mention that
at zero temperature the
asymptotic behavior of the prekernel can be obtained directly from
its definition (\ref{prekernel}) and agrees with (\ref{surfacecor}).

On the other hand, 
in the high temperature limit
the Fermi-Dirac distribution in
(\ref{prekernel})
can be replaced by a Boltzmann distribution.
The prekernel is thus given
by
\be
k(z_1, z_2) = \e^{-z_1 z_2^*} \sum_{k=0}^\infty \frac{(z_1 z_2^*)^k}
{k!}\zeta\,\e^{-\beta k}.
\label{prekernel1}
\ee
In this limit the fugacity is equal to $\zeta = \beta n$,
resulting in
\be
k(z_1,z_2) =\beta n.
\ee
This requires us to define the scaled temperature by
\be
\beta=\frac{1}{h n},
\ee
as opposed to the low-temperature case (\ref{hn}).
The spectral density is thus given by
\be
 \rho(z) = \frac{1}{\pi h},
\ee
and the two-point correlation function has the
exponential
form
\be
R_2(z_1,z_2)=- \frac{1}{\pi^2 h^2}\e^{-|z_1-z_2|^2} .
\ee
Since the average spectral density
decreases as $1/h$, the unfolded eigenvalues
become uncorrelated (Poisson statistics) in the high temperature limit.

\section{Weak nonhermiticity}
In the case of weak nonhermiticity,
we start from the identity
\be
\e^{\frac{\alpha}{\tau (1-\alpha^2)} z_i z_j^* } &=& \sqrt{1-\alpha^2}
\e^{\frac{\alpha^2 }{2\tau(1-\alpha^2)}(z_i^2+z_j^{*\,^2 })}
\sum_{m=0}^\infty
\frac{\alpha^m}{m!} H_m(\frac{z_i}{\sqrt \tau})
H_m(\frac{z_j^*}{\sqrt \tau}), \label{iweak}
\ee
where $H_m(z)$ are the
Hermite polynomials.
Performing  exactly the same manipulations as in (\ref{manipul}) we
obtain
\be
\det \e^{\frac \alpha{\tau (1-\alpha^2)}z_i z_j^*}
&=&(\sqrt{1-\alpha^2})^n \sum_{m_1< m_2< \cdots < m_n}
\alpha^{m_1+\cdots +m_n} \nn \\
&&\times
\det \frac{\e^{\frac{\alpha^2}{2\tau(1-\alpha^2)} z_j^2}
H_{m_i}(\frac{z_j}{\sqrt\tau})}
{\sqrt{m_i!}}
\det \frac{\e^{\frac{\alpha}{2\tau (1-\alpha^2)} z_k^{*2}}
H_{m_l}(\frac{z_k^*}{\sqrt \tau })}{\sqrt{m_l!}}.
\ee
The joint probability distribution (\ref{our-joint}) can
thus be written as
\be
P(z)\sim
\pi^n (1-\alpha^2)^{n/2}(1-\tau^2)^{n/2} \sum_{m_1< m_2< \cdots < m_n}
\left (\frac \alpha {\tau}\right )^{m_1+\cdots +m_n}
\det \phi_{m_i}(z_j) \det \phi_{m_k}(z^*_l),
\label{jointweak}
\ee
where the  wave functions defined by
\be
\phi_k(z) = \frac{{\tau}^{k/2}}{\sqrt \pi (1-{\tau}^2)^{1/4} \sqrt{k!}}
H_{k}(\frac z{\sqrt {\tau} })\, \e^{- \frac 12\frac {1}{1-\tau^2}[|z|^2
- \tau   z^2 ] }
\label{waveweak}
\ee
satisfy the orthogonality relations \cite{difrancesco}
\be
\int d^2z\,\phi_k(z^*)\phi_l(z) =
\delta_{kl} .
\ee
The above wave functions (\ref{waveweak}) also span the
set of the single particle wave functions
in the lowest Landau level
obeying the Schr\"{o}dinger equation
(\ref{hamiltonian}-\ref{schrodinger}) which, in terms of properly
rescaled coordinates, reads
\be
\left [\frac 12 (1-\tau^2)(i\partial_x -\frac y{1-\tau^2})^2 +
\frac 12 (1-\tau^2)(i\partial_y +\frac x{1-\tau^2})^2\right ]
\phi_m = \phi_m.
\label{hamt}
\ee
If we write
\be
\frac \alpha \tau = \e^{-\beta},
\ee
the joint eigenvalue distribution
may be interpreted as the diagonal element of
the $n$-body density matrix
of the lowest Landau level fermions at temperature $1/\beta$. 
The Schr\"odinger equation corresponding to (\ref{schrm}) now reads
\be
\left [\frac 12 (1+\tau)(i\partial_x +\frac y{1-\tau^2})^2 +
\frac 12 (1-\tau)(i\partial_y -\frac x{1-\tau^2})^2
+\frac \tau{1-\tau^2}(x+iy)^2 \right ]
\phi_m = (2m+1)\phi_m . \nonumber \\
\label{hamt2}
\ee
Although,  this relation is physically appealing we do not rely on it to
obtain our results.

Now we turn to the calculation of correlation functions.
The $p$-particle  correlation function is obtained by integrating
$P(z_1, \cdots, z_n)$ over
$z_{p+1}, \cdots, z_n$. Using the orthogonality of the wave functions
and
expressing (\ref{jointweak}) as a single determinant
one easily finds
\be
R_p^n(z_1, \cdots, z_p) &=& \frac{n!}{(n-p)!}
\int d^2 z_{p+1} \cdots d^2 z_n P_n(z)\nn \\
&=&\frac{1}{Z_n}\sum_{m_1< m_2< \cdots < m_n}
\det_{i,j=1,\ldots,p} \sum_{k=1}^n \e^{-\beta m_k}
\phi_{m_k}(z_i) \phi_{m_k}(z^*_j).
\ee
Here, the overall normalization constants $Z_{n}$ have been chosen such
that
the joint probability integrates to unity.
In an occupation number representation this correlator can be
written as
\be
R_p^n(z_1, \cdots, z_p) = \frac{1}{Z_n}
\sum_{n_0+ n_1 +\cdots  = n}
 \det_{i,j=1,\ldots,p}
\e^{-\beta\sum_q q n_q}
\sum_{k=0}^\infty  n_k
\phi_{k}(z_i) \phi_{k}(z^*_j),
\ee
where the occupation number $n_k$ runs over $\{0,1\}$. Such sums are
easily calculated in the grand canonical ensemble
\be
R_p(z_1, \cdots, z_p) &=&
\frac 1Z \sum_{n=0}^\infty \zeta^n Z_n R_p^n(z_1,\cdots,z_p),
\ee
where $\zeta$ is the fugacity and
$Z$ is the grand canonical partition function given by
\be
Z = \prod_{k=0}^\infty (1+\zeta \e^{-\beta k}).
\ee

In the thermodynamic limit
the correlators obtained by means of the grand canonical ensemble
coincide with those from the canonical ensemble.
The sum of the $n_k$ can now be performed easily. The result is given
by
\be
R_p(z_1,\cdots, z_p) = \det_{i,j=1,\ldots,p}  K(z_i,z_j),
\ee
with kernel defined by
\be
K(z_i,z_j) =
\sum_{k=0}^\infty
\frac{\phi_{k}(z_i) \phi_{k}(z^*_j)}{1+\zeta^{-1} \e^{\beta k}}.
\label{kerneldef}
\ee
The average spectral density,  obtained by integrating over all
eigenvalues
except one, is thus given by
\be
\rho(z) = K(z,z)
= \sum_{k=0}^\infty
\frac{\phi_{k}(z) \phi_{k}(z^*)}{1+\zeta^{-1} \e^{\beta k}} .
\ee
The fugacity follows from the normalization integral and is given by
\be
n = \sum_{k=0}^\infty
\frac{1}{1+\zeta^{-1} \e^{\beta k}}.
\ee
Similarly,
the two-point correlation function is obtained by integrating over all
eigenvalues except two.
Subtracting $ \rho(z_1) \rho(z_2)$  results is the
connected two-point correlation function given by
\be
R_2(z_1,z_2) = - |K(z_1, z_2)|^2.
\ee

As in the case of strong nonhermiticity, the kernel can be simplified
by means of a partial resummation
\be
K(z_i,z_j) &=&
\sum_{m=0}^\infty  \sum_{k=0}^m
\phi_{k}(z_i) \phi_{k}^*(z_j)
\left [\frac{1}{1+\zeta^{-1} \e^{\beta m}}-
\frac{1}{1+\zeta^{-1} \e^{\beta (m+1)}} \right ]\nn \\
&=&
\sum_{m=0}^\infty  K_m^0(z_i,z_j)
\left [\frac{1}{1+\zeta^{-1} \e^{\beta m}}-
\frac{1}{1+\zeta^{-1} \e^{\beta (m+1)}} \right ],
\label{kernel-def}
\ee
where the zero temperature kernel is defined by
\be
K_m^0(z_i,z_j) =
\sum_{k=0}^m
\phi_{k}(z_i) \phi_{k}(z^*_j).
\label{K0}
\ee

\subsection{Correlations in the bulk}
The bulk scaling limit of the zero temperature kernel (\ref{K0})
was analyzed in detail in \cite{Fyodorov2}.
We will recall their method for the sake of completeness.
Using an integral representation of the Hermite polynomials,
it can be  rewritten as
\be
K_m^0(z_1,z_2) &=&
\frac{1}{2\pi^2 \tau \sqrt{1-\tau^2}}
\e^{-\frac{1}{2(1-\tau^2)}[ |z_1|^2+|z_2^2| - \frac \tau 2 (z_1^2
+z_2^2+z_1^{*2}+z_2^{*2}) ] +\frac{1}{2\tau}(z_1^2 +z_2^{*2})}
\nonumber\\
&&\times
\int_{-\infty}^\infty \int_{-\infty}^\infty
{dr\, ds}\,
{\e^{(- r^2/2 +i r z_1 -s^2/2 -i s z^*_2)/\tau+ r s}}
\frac{\Gamma(m+1, rs)}{m!}
\label{Kint}\\
&=&
\frac{1}{\pi^2 \tau \sqrt{1-\tau^2}}
\e^{-\frac{1}{2(1-\tau^2)}[ |z_1|^2+|z_2^2| - \frac \tau 2 (z_1^2
+z_2^2+z_1^{*2}+z_2^{*2}) ] +\frac{1}{2\tau}(z_1^2 +z_2^{*2})}
\nn\\
&&\times
\int_{-\infty}^\infty\!
\int_{-\infty}^\infty
\!\!
du\, dv\, \e^{u^2(1-1/\tau)-v^2(1+1/\tau) +iu(z_1-z_2^*)/\tau
+iv(z_1+z_2^*)/\tau}  \frac{\Gamma(m+1, u^2-v^2)}{m!} .
\nn
\ee
where $r=u+v$ and $s=u-v$.
The $v$-integral can be performed by a saddle-point approximation.
To the leading order,
the argument $v$ in the incomplete $\Gamma$-function can
be replaced by its saddle-point value given by
\be
\bar v = \frac {i(z_1+z_2^*)}{2(1+\tau)} .
\ee
For $u^2-v^2 \sim m$ and $m \to \infty$,
the incomplete $\Gamma$-function can be approximated by
a step function
\be
\frac{1}{m!}\Gamma(m+1, u^2-v^2) \approx 1  \quad {\rm for} \quad
u^2 < m + \bar v^2 = m - \frac {x^2}{(1+\tau)^2}
\label{gamma-approx}
\ee
and zero otherwise, depending on whether its integration domain
contains the saddle point or not.
We thus find the kernel
\be
K_m^0(z_1,z_2) &=&
\frac{\sqrt \pi }{\pi^2 \tau \sqrt{1-\tau^2}\sqrt{1+1/\tau}}
\int_{-\sqrt{m+\bar v^2}}^{\sqrt{m+\bar v^2}}
 du \, \e^{u^2(1-1/\tau) +iu(z_1-z_2^*)/\tau }  \nn\\
&&\times \e^{-\frac{1}{2(1-\tau^2)}[ |z_1|^2+|z_2^2| -
\frac \tau 2 (z_1^2
+z_2^2+z_1^{*2}+z_2^{*2}) ] +\frac{1}{2\tau}(z_1^2 +z_2^{*2})
-\frac{ (z_1+z_2^*)^2}{4\tau(\tau+1)}}.
\ee
In the limit of weak nonhermiticity 
we magnify the bulk of the spectrum according
\be
\label{para}
z_1 &=& x\sqrt n +\frac{\pi r}{2\sqrt n} +i\frac{y_1}{\sqrt n},
\nn\\
z_2 &=& x\sqrt n -\frac{\pi r}{2\sqrt n} +i\frac{y_2}{\sqrt n},\nn\\
\tau^2 &=& 1 - \frac {a^2}n,
\ee
where $-2<x<2$. For $n \to \infty$ this results in
\be
\label{zerotempK}
K_m^0(z_1,z_2) &=&
\frac{n\sqrt {\pi }}{\pi^2 a\sqrt{2} }
\e^{-\frac{1}{a^2}(y_1^2 +y_2^2) +\frac i2 x(y_1-y_2)}
\int_{-\sqrt{(m+\bar v^2)/n}}^{\sqrt{(m+\bar v^2)/n}}
 du\,  \e^{-\frac {a^2 u^2}2 +iu(\pi r+i(y_1+y_2)) } .
\ee

For $\beta \to 0$ the sum over $m$ can be replaced by an integral.
In this limit the kernel (\ref{kernel-def}) is given by
\be
K(z_1, z_2) &=& \int_{-1+x^2/4}^\infty ndt \frac {\beta
K_{n(1+t)}^0(z_1,z_2)}{4\cosh^2(\beta n t /2)}\nn \\
&=& \frac{n^2}{\pi a\sqrt {2\pi}}\int_{-1+x^2/4}^\infty dt
\frac {2\beta } {4\cosh^2(\beta n t /2)}
\int_{0}^{\sqrt{1+t-x^2/4}}
 du  e^{-\frac {a^2 u^2}2} \cos u(\pi r+i(y_1+y_2))  \nn\\
&&\times e^{-\frac 1{a^2}(y_1^2 +y_2^2) +\frac i2 x(y_1-y_2) }\nn \\
&=&
\frac{n}{\pi a\sqrt {2\pi}}\int_{-\infty}^\infty dp
\frac {1} {1+e^{(p^2-1+x^2/4)/h}}
e^{-\frac {a^2 p^2}2} e^{ip(\pi r+i(y_1+y_2))}
e^{-\frac 1{a^2}(y_1^2 +y_2^2) +\frac i2 x(y_1-y_2) }\nn \\
\label{copon}
\ee
where the combination
\be
\label{bulkh}
n\beta \equiv \frac 1h
\ee
is kept fixed in the thermodynamic limit.
Finally,
we derive the small $h$ limit of the kernel for $x$ in
 the center of the spectrum  $(x \approx 0)$.
The second integral in (\ref{copon}) is  rewritten
by expressing the Gaussian term as,
\be
\label{hub}
e^{-\frac {a^2u^2}2 +i u (\pi r+i(y_1+y_2))} = \frac 1{a\sqrt{2\pi}}
\int_{-\infty}^\infty
ds e^{-\frac{(s - \pi r-i(y_1+y_2))^2}{2 a^2} + isu}
\ee
After performing the integral over $u$ we obtain
\be
K(z_1, z_2) =
\frac{ n}{ \pi^2  a^2 }
\int_{-\infty}^\infty  ds e^{\frac{-(s-\pi r-i(y_1+y_2))^2}{2a^2}}
\int_{-1}^\infty
dt \frac {\sin(s\sqrt{1 + t})}{\cosh^2\frac t{2h}}
e^{-\frac 1{a^2}(y_1^2 +y_2^2)     }
\ee
The integral over $t$ can be performed to leading order  in $h$. In that
case $\sqrt{1 + t}$ can be expanded to first order in $t$
and  the resulting integral over $t$, after extending its
lower limit to $-\infty$, is known analytically.
We finally obtain
\be
K(z_1, z_2) =
\frac{ n h  }
{2 \pi  a^2 } e^{-\frac {y_1^2 +y_2^2}{a^2}  }
\int_{-\infty}^\infty  ds e^{\frac{-(s-\pi r-i(y_1+y_2))^2}{2a^2}}
\frac {\sin s  }{\sinh(\pi sh /2)}.
\label{smallhkernel}
\ee
Sometimes it is useful to explicitly display the $h=0$ contribution
to the kernel.  From the
second integral in (\ref{copon}) at $h=0$
one can explicitly find the zero temperature
 result reported in {\cite{Fyodorov1}}. By subtracting and
 adding this term to  (\ref{smallhkernel}) we find
\be
\label{hoste}
K(z_1,z_{2})=\frac{2n}{\pi a
}\frac{1}{\sqrt{2\pi}}e^{-\frac{y_{1}^2+y_{2}^2}{a^{2}}}\left[
\int_{0}^{1}due^{-\frac{(au)^2}{2}}\cos(u(\pi r+i(y_{1}+y_{2})))+
\frac{\pi h}{2a\sqrt{2\pi}}\right. \\ \nonumber
\left.\int_{-\infty}^{\infty}ds \left(\frac{\sin s}
{\sinh(\pi h s/2)}-
\frac{\sin s}{\pi hs/2}\right)e^{-\frac{1}{2a^{2}}(s-(r\pi+
i(y_{1}+y_{2})))^2}\right],
\ee
where the first and third integrals cancel each other.

\begin{figure}[t]
\begin{center}
\epsfig{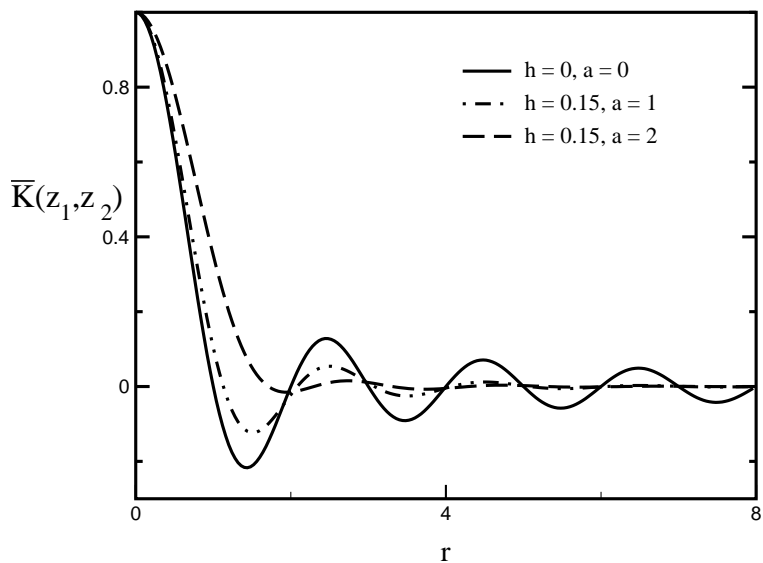}
\caption[]{${\bar K}(z_{1},z_{2})$ (\ref{rker}) at $x=y_{1}=y_{2}=0$.
}
\end{center}
\label{fig1}
\end{figure}
The spectral density at the center of the band is given by,
\be
\label{once}
 \rho(y) = K(z=iy/{\sqrt{n}},z=iy/{\sqrt{n}})=\frac{2n}{\pi
a}\frac{1}{\sqrt{2\pi}}\left[e^{-\frac{2y^2}{a^{2}}}
\int_{0}^{1}dte^{-\frac{a^{2}t^2}{2}}\cosh(2ty)+
\frac{\pi h}{a\sqrt{2\pi}}\right. \\ \nonumber
\left.\int_{0}^{\infty}dt \left(\frac{\sin t}
{\sinh(\pi h t/2)}-
\frac{\sin t}{\pi ht/2}\right)e^{-\frac{t^2}{2a^{2}}}\cos(2yt/a^{2})\right]
\ee
where $y_1=y_2=y$.
The integral over ${\rm Im}(z)$ of the spectral density is given by
\be
\int_{-\infty}^\infty K(z,z)d{\rm Im}z  &=&
\frac 1{\sqrt n} \int_{-\infty}^\infty \rho(y)  dy\nn \\
&=& \frac{\sqrt n} \pi
\label{densx}
\ee

In Fig. 1, we show  the normalized kernel ${\bar K}(z_1,z_2)$
defined by
\be
\label{rker}
{\bar K}(z_{1},{z_{2}})=\frac{K(z_{1},{z_{2}})}
{\sqrt{ \rho(z_{1})\rho(z_2)}}
\ee
for $h =0.15$ and different
values of the nonhermiticity parameter.
 We find that the spectral
correlations weaken for  increasing values of $a$ and approach the
result for the  Ginibre ensemble for
$a \approx 2$.
 Although not shown in the picture, it was verified numerically that
 the exact result (\ref{copon})  is
almost indistinguishable  from the small $h$ result (\ref{hoste})
for values of $h$ up to $h \sim 0.3$, and significant differences
are only found for values of $h$ as large as
 $h \approx 1$.
The normalized critical kernel for Hermitian ensembles \cite{MSN}
is easily reproduced from the ratio (\ref{rker})
starting from the expression (\ref{smallhkernel}) and taking the
limit $a \to 0$,
\be
\label{op}
\bar K(z_1,z_2)\to \frac{\pi h}{2}\frac{\sin(\pi r)}{\sinh(\pi^2 r h /2)}.
\ee
If we consider the $a \to 0$ limit of the kernel (\ref{hoste})
 or the spectral density (\ref{once}),
$\delta$-functions of the imaginary part of the eigenvalues have
to be taken into account carefully. For example, the $a \to 0$ limit of the
spectral density (\ref{once}) is given by
\be
\rho(z) = \frac n \pi \delta(y).
\ee

Finally, let us mention that for $a \gg 1$
we recover the Ginibre's kernel for
general complex matrices.


\subsection{Correlations at the edge}

Next we consider a microscopic scaling limit
at the vicinity of either edge of the band of eigenvalues for
$z\sim \pm 2\sqrt{n}$,
as an extension of edge correlation of
the Hermitian Random Matrix ensembles.

We shall need a more refined asymptotic formula for the
incomplete $\Gamma$-function than Eq. (\ref{gamma-approx}).
For $x{> \atop \sim}m$ and $m\gg 1$,
the incomplete $\Gamma$-function is dominated by the
contribution from the lower end point,
so that \cite{choquard,forrester}
\be
\Gamma(m+1,x)=\e^{-x}\frac{x^{m+1}}{x-m}
\left[
1+O\left(\frac{m}{(x-m)^2}\right) \right] .
\label{asympt}
\ee
Accordingly, the kernel at zero temperature
(\ref{Kint}) reads
\be
K_m^0(z_1,z_2) &\simeq&
\frac{1}{2\pi^2 \tau \sqrt{1-\tau^2}}
\e^{-\frac{1}{2(1-\tau^2)}[ |z_1|^2+|z_2^2| - \frac \tau 2 (z_1^2
+z_2^2+z_1^{*2}+z_2^{*2}) ] +\frac{1}{2\tau}(z_1^2 +z_2^{*2})}
\nonumber\\
&&\times
\int_{-\infty}^\infty \int_{-\infty}^\infty \frac{dr\, ds}{rs-m}
\e^{(- r^2/2 +i r z_1 -s^2/2 -i s z^*_2)/\tau
+ (m+1)\log rs-\log m!} .
\ee
For $z_1,z_2 \sim 2\sqrt{n}$, $m\sim n$, and $\tau\sim 1$,
the two saddle points of the $r$ ($s$) integral
merge at $r=i\sqrt{n}$ ($s=-i\sqrt{n}$).
In order to obtain a nontrivial result,
we magnify this region according to the scaling
\be
z_i&=&2\sqrt{n}+\frac{x_i}{n^{1/6}}+i \frac{y_i}{n^{1/2}},
\nn\\
m&=&n+{n^{1/3}}t ,
\label{mnt}
\\
\tau^2 &=& 1-\frac{\alpha^2}{n},
\nn
\ee
and change the integration variables as
\be
r=i\sqrt{n}+ n^{1/6} p,\quad s=-i\sqrt{n}- n^{1/6} q.
\ee
The subleading terms in (\ref{asympt}) are of order
$O(n^{-1/6})$ in this scaling limit and can be ignored.
To the leading order in $n$ we obtain
\be
K_m^0(z_1,z_2)&=&
\sqrt{\frac{2}{\pi}}\frac{n^{1/3}}{a}
\e^{i(y_1-y_2) - \frac{1}{a^2} (y_1^2 +y_2^2)}
 \int_{-\infty}^{\infty}  \int_{-\infty}^{\infty}
 \frac{dp}{2\pi}\frac{dq}{2\pi}
 \frac{\e^{
   i\frac{p^3}{3} + i p{(x_1-t)}
 + i\frac{q^3}{3} + i q{(x_2-t)}}}{-i(p + q)}
\nn\\
&=&\sqrt{\frac{2}{\pi}}\frac{n^{1/3}}{a}
\e^{i(y_1-y_2) - \frac{1}{a^2} (y_1^2 +y_2^2)}
\int_{-\infty}^{t}\! dt'\,{\rm Ai}(x_1-t'){\rm Ai}(x_2-t').
\label{airyt}
\ee
where ${\rm Ai}(x)$ is the Airy function
\be
{\rm Ai}(x)=
\int_{{-\infty}}^{{\infty}} \frac{dp}{2\pi}
\e^{i \frac{p^3}{3} + i p x}=
\int_{0}^{{\infty}} \frac{dp}{\pi}
\cos\Bigl(\frac{p^3}{3} +  p x\Bigr) .
\ee
The integral in Eq.\ (\ref{airyt})
is called the Airy kernel $K_{\rm Ai}(x_1-t,x_2-t)$
(See Ref.\cite{Dyson-Mehtabook}, \S{}18), describing the
edge correlations of the Gaussian Unitary Ensemble. By partial 
integrations one may express it in an alternative and more familiar form
\be
K_{{\rm Ai}}({x_1},{x_2})=
\frac{{\rm Ai}(x_1){\rm Ai}'(x_2)-
{\rm Ai}'(x_1){\rm Ai}(x_2)}{x_1-x_2}.
\ee

The scaling  of $m$ in (\ref{mnt}) requires the introduction
of a finite temperature parameter $h$ by
\be
\beta=\frac{1}{n^{1/3} h},
\ee
in contrast to the
bulk scaling (\ref{bulkh}).
After replacing the sum over $m$ by an integral over $t$,
the low-temperature limit of the kernel
(\ref{kernel-def}) is given by
\be
K(z_1,z_2)&=&
\sqrt{\frac{2}{\pi}}\frac{n^{1/3}}{a}
\e^{i(y_1-y_2) - \frac{1}{a^2} (y_1^2 +y_2^2)}
\int_{-n^{2/3}}^\infty dt\,
K_{{\rm Ai}}(x_1-t,x_2-t)
\frac{d}{dt}\left(\frac{1}{1+e^{t/h}}\right)
 \nn\\
&=&\sqrt{\frac{2}{\pi}}\frac{n^{1/3}}{a}
\e^{i(y_1-y_2) - \frac{1}{a^2} (y_1^2 +y_2^2)}
\int_{-\infty}^{\infty} dt
\frac{{\rm Ai}(x_1-t){\rm Ai}(x_2-t)}{1+e^{t/h}}.
\ee
Due to the different orders of the
level spacings in real and imaginary directions,
the zero-temperature kernel is factorized,
unlike the bulk kernel Eq. (52) of \cite{Fyodorov}
or our Eq. (\ref{zerotempK}).
Namely, the dependence of $K^0_m$ on the order $m$
is merely to dilate the eigenvalue support, which
can be compensated by a change of the real part of
the eigenvalue coordinate, $x\to x-t$.
Accordingly, the effects of nonhermiticity and
finite temperature are factorized.
The former is
reflected in the scaled kernel as a Gaussian blurring
in the $y$-direction whereas, as the temperature $h$ increases,
the oscillation of the scaled spectral density
along the $x$-direction is weakened
toward the Poissonian limit. 
This is shown in Fig. 2 where we plot the spectral density in the Hermitian
limit given by
$\rho(x)=\int dt\,{\rm
Ai}(x-t)^2/(1+\e^{t/h})$.
\begin{figure}[!tpb]
\begin{center}
\epsfig{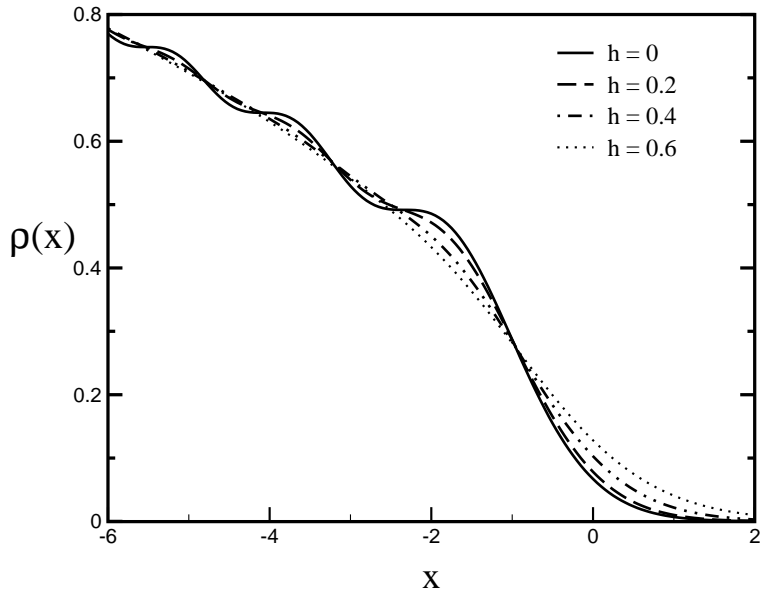}
\caption[]{The spectral density at the edge
for different values of the temperature parameter,
at zero nonhermiticity.}
\end{center}
\label{fig2}
\end{figure}


\section{Number variance}

The number variance in an arbitrary domain $A$ of the complex plane
is given by,
\be
\label{deus}
\Sigma_{2}(L)~=~L-\int_{A}d^{2}z_{1}
\int_{A}d^{2}z_{2}Y_{2}(z_{1},z_{2}) \qquad {\rm with} \qquad
L = \int_A d^2 z \rho(z),
\ee
where $\rho(z)=K(z,z)$,
$Y_{2}(z_{1},z_{2})=|K(z_{1},z_{2})|^2$ and
$K(z_{1},z_{2})$ is the spectral kernel defined in (\ref{copon}).
Apart from edge correlations we have found that in the strong
nonhermiticity case the two-point correlations decay exponentially
on a scale of one level spacing or less which results in an asymptotic
linear dependence of the number variance on $A$ with unit slope.
Below we focus our
analysis on the more interesting weak nonhermiticity limit.

As will be seen in the figures below, the
fluctuations of the eigenvalues  increase  with both increasing
temperature $h$
and increasing degree of week nonhermiticity  $a$. The reasons
for such behavior are the following: For larger values of $h$,
the correlations of distant eigenvalues
are suppressed resulting in stronger fluctuations and  the slope
of the asymptotically linear number variance increases with $h$.
By increasing the degree of nonhermiticity,
eigenvalues have more room to avoid each other along the imaginary axis.
As a consequence, spectral fluctuations are stronger and deviations from
Wigner statistics are observed.

\begin{figure}[!tpb]
\begin{center}
\epsfig{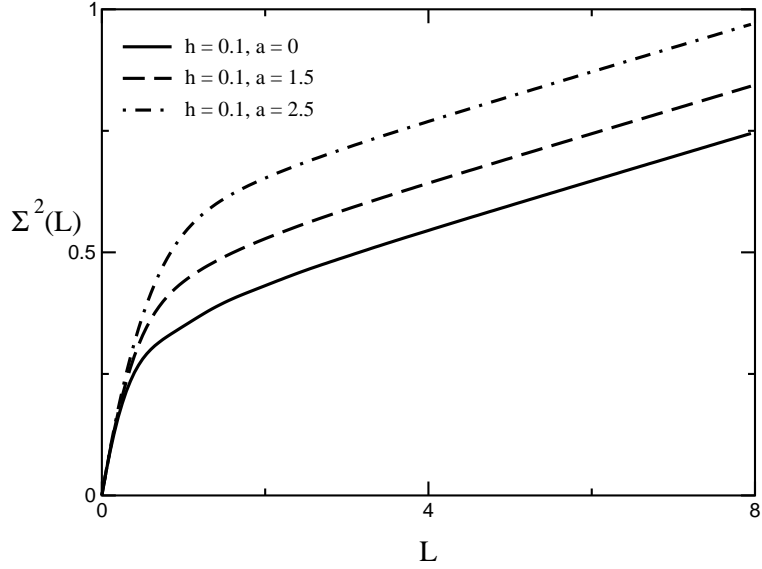}
\caption[]{The small $h$ behavior of the number variance $\Sigma^2(L)$
versus $L$
given in (\ref{diez}) for $h=0.1$ and values
of the nonhermiticity parameter as given in the legend of the figure.
}
\end{center}
\label{fig3}
\end{figure}

In the limit  $h \ll 1 $ we calculate the number variance for the area
$A=[-L_x/2,L_x/2] \times ( -\infty , \infty )$. Because of
the normalization integral (\ref{densx}) we choose $L_x = L \pi/\sqrt n$
so that the area $A$ contains $L$ eigenvalues on average. The dependence
of the kernel on $x$ is subleading in
the thermodynamic limit. This allows us to rewrite the number variance as
\be
\Sigma^2(L)=L-\frac {2\pi }{n^{3/2}} \int_{0}^{L}
dr(\pi L/\sqrt{n}- \pi r /\sqrt n)
\int_{-\infty}^{\infty}\int_{-\infty}^\infty dy_1dy_2
| K(z_1,z_2)|^2,
\ee
where the prefactor includes a contribution from the Jacobian of
the transformation (\ref{para}). The
 integrals over $y_1$ and $y_2$ are easily performed in terms of
the variables $u \equiv y_1+y_2$ and $v\equiv y_1-y_2$. The final result
for the small $h$ limit of the number variance is thus given by
\be
\label{diez}
\Sigma^{2}(L)=L-2\int_{0}^{L}dr(L-r)\left[\frac{\sin^{2}(\pi
r)}{{\pi}^{2} r^{2}}e^{-\frac{a^{2}r^{2}}{L^{2}}}\right . +  \nonumber\\
\left .\frac{\pi^2
h^2}{4}\frac{1}{a\sqrt{\pi}}\int_{-\infty}^{\infty}dt\left(
\frac{\sin^{2}(t)}{\sinh^{2}(\pi ht/2)}-\frac{\sin^{2}(t)}{(\pi
ht/2)^{2}}\right)
e^{-\frac{1}{a^2}(t-\pi r)^{2}}\right] .
\ee
\begin{figure}[!ht]
\begin{center}
\epsfig{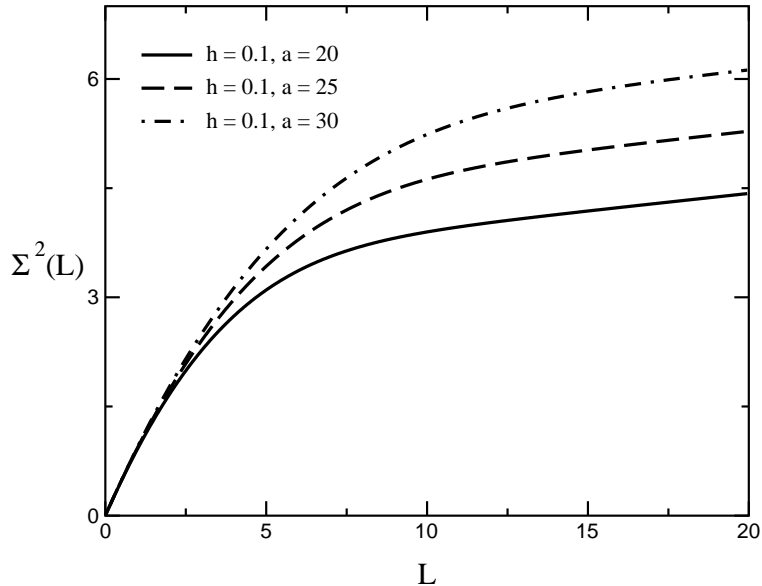}
\caption[]{The number variance (\ref{diez}) is computed
  for large values of the nonhermiticity parameter $a$.}
\end{center}
\label{fig4}
\end{figure}
We observe that in this limit the
finite temperature effects decouple from the weak nonhermiticity
corrections. For $L \gg 1/h$ and $a \ll L$ it can be shown from
 (\ref{diez}) that the number variance
is given by
\be
\Sigma^{2}(L)= \frac a{\pi^{3/2}} - \frac \gamma{\pi^2}
+\frac h2 L + O(1/L),
\label{asymnum}
\ee
where $\gamma$ is the Euler constant. The term linear in $a$ can
be calculated in the $h \to 0$ limit and was obtained in \cite{Fyodorov},
whereas the term linear in $h$ can be calculated for $a \to 0$ and
was derived in \cite{MSN}.
In Fig. 3, we show the small $h$ limit of the number variance
(\ref{diez}) for $h=0.1$
and different values of the
nonhermiticity parameter. We observe that the asymptotic linear
behavior given by (\ref{asymnum}) is already reached well below
the expected scale of $1/h$.
We remark that for values of $h$ as large as $0.3$ the small
$h$ result (\ref{deus}) is still very close to the exact result
obtained with the kernel  (\ref{copon}).

\begin{figure}[!ht]
\begin{center}
\epsfig{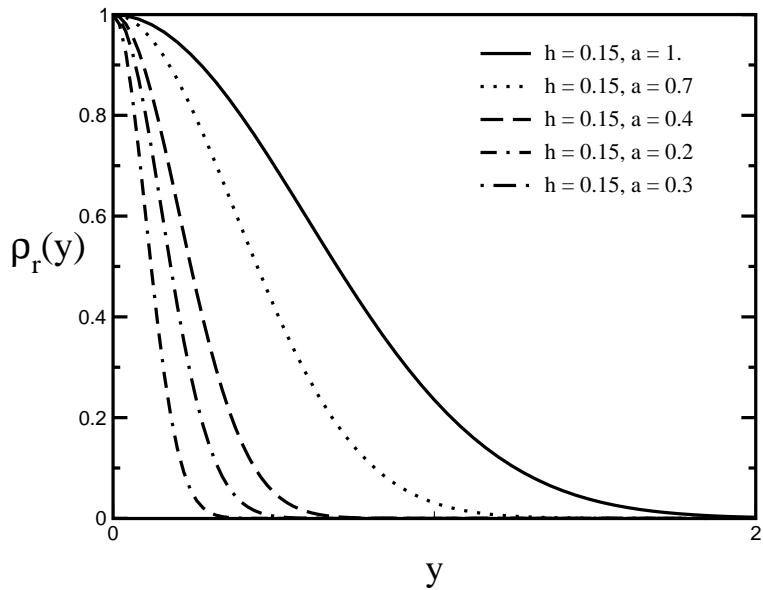}
\caption[]{The renormalized
 spectral density $\rho_{r}(y)= \rho(y)/\rho(0)$ (with $\rho(y)$ defined in
(\ref{once}) 
in the center of the band
is shown for  different values of the nonhermiticity parameter. }
\end{center}
\label{fig5}
\end{figure}

The small $h$ result for the number variance
(\ref{diez}) is also valid for large values of the
nonhermiticity parameter. Plots of (\ref{diez}) for $a \gg 1$ are
shown in Fig. 4. We find that the asymptotic result for the slope
is still approximately given by $h/2$ and
depends only weakly on $a$.
For $L \ll a$ we find that
$\Sigma^2(L) \to L$  which is the result for strong
nonhermiticity. This
crossover behavior
was first found in the limit $h \to 0$ \cite{Fyodorov}.

The imaginary part of the eigenvalues is of order $a$. This is
shown in Fig. 5 where we plot the
$\rho(y)/ \rho(0) $ (with $\rho(y)$ given in eq.
(\ref{once})) versus $y$.
Since the imaginary part of the eigenvalues is of the same
order as the spacing of the real part of the eigenvalues, the
number variance computed for a rectangle
$0<{\rm Im} z<\Delta y \ll a$ is expected to be given by $\Sigma^2(L) \to
L$
where $L$ is the total number of eigenvalues in the rectangle.
This is shown in Fig. 5 where we plot the number
variance obtained from (\ref{deus}) using the kernel (\ref{hoste}).


\section{Conclusions}

In this article we have introduced
a two parameter  ensemble of complex random matrices
with no hermiticity conditions imposed. This
ensemble  interpolates  between  
 the Gaussian Unitary Ensemble, the Ginibre ensemble and the Poisson ensemble.
Using methods from statistical mechanics and properties of orthogonal
polynomials, we have analyzed this ensemble in two different limits:
weak nonhermiticity and strong nonhermiticity.

We have shown that the joint eigenvalue distribution of our random matrix
model coincides with the diagonal element of the density matrix
 of  a two   dimensional gas of spinless fermions in the
lowest Landau  level at finite temperature.  The two parameters
of our model have been interpreted in terms of a shape parameter
of the two dimensional  domain of eigenvalues (or particles)
and a temperature.

\begin{figure}[!ht]
\begin{center}
{\epsfig{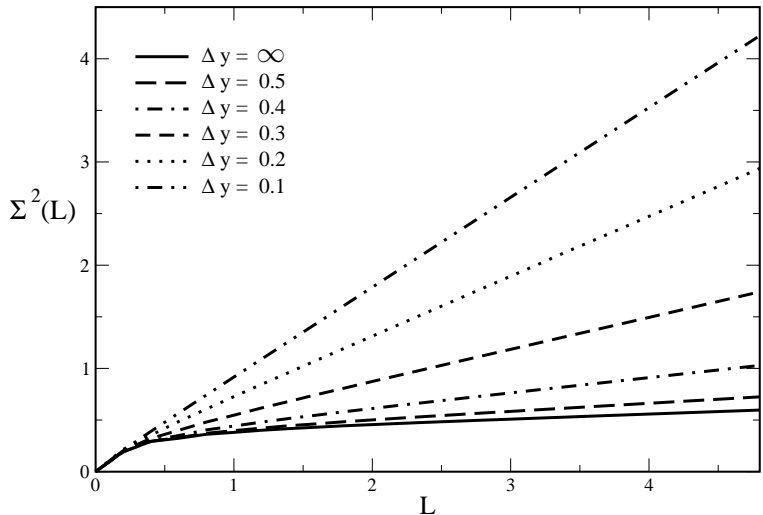}
\setlength{\unitlength}{1.0in}
\begin{picture}(1,1)
\put(-2.75,2.37){\small $\infty$}
\end{picture} }
\caption[]{The number variance given by the general
formula (\ref{deus}). The domain of integration
is a rectangle 
in the complex plane containing $L$ eigenvalues and width given by
$0<{\rm Im } z < \Delta y$.  The nonhermiticity parameter is equal to
 $a=0.4$  and the  value of $h$ is equal $h=0.1$
for all curves.   The number
variance is almost Poisson
for $\Delta y \leq a$.
}
\end{center}
\label{fig6}
\end{figure}

In the strong nonhermiticity limit, in the bulk of
the spectrum, the correlations of the eigenvalues are given by
Ginibre statistics and decrease exponentially on the scale of
the average level spacing.
The situation is different near the surface of the spectrum where,
at zero temperature,
the correlations decrease as an inverse square law in the direction
of the surface. At finite temperature this power-law behavior
changes into an exponential behavior.
At very high temperatures the surface and the bulk are no longer
distinguishable. In that case the two-point correlation function
of the unfolded eigenvalues
still decays exponentially but with an exponent that is proportional
to the temperature. In this way  the Poisson limit is recovered at
high temperatures.

In the weak nonhermiticity limit there is no clear distinction between
bulk and surface and the temperature affects the correlation functions
of  the eigenvalues.
In the low temperature limit we have obtained
a closed analytical
expression for the two-point correlation function which reproduces critical
 statistics.
We have found that, although level repulsion is still present,
the number variance is asymptotically linear with a slope depending
on the temperature parameter but not on the nonhermiticity parameter.
A remarkable feature  is that
temperature and weak nonhermiticity effects decouple in this region.
Thus critical statistics is not modified by a
 weak nonhermitian perturbation.

 Finally, let us explain a physical prediction of the present model. Since
for critical statistics
 the slope of the number variance is related to
 the multifractal dimension of the wave function and, in our model,
the slope does not depend on the nonhermiticity parameter,
we predict that
 the multifractal dimension of a physical system does not depend on
the nonhermiticity parameter either. We thus predict the same multifractal
dimensions for open and dissipative systems. A simple model for which this
 prediction may be tested is a three dimensional
disordered system at the critical density of impurities and
with several leads attached to it.  We thus expect that in the
weak nonhermiticity domain the leads do not affect
the multifractal dimension of the wavefunctions.


\vskip 1.5cm
\noindent
{\large {\bf Acknowledgments}}\\

This work is was partially supported  by
the Department of Energy grants No.
DE-FG02-92ER40716 (SMN) and DE-FG-88ER40388 (AMG and JJMV).

\end{document}